\documentclass[12pt]{article}

\usepackage{graphicx}

\begin{document}

\input{epsf.sty}

\begin{titlepage}

\begin{flushright}
IUHET-505\\
hep-th/0701270
\end{flushright}
\vskip 2.5cm

\begin{center}
{\Large \bf Cerenkov Radiation in a Lorentz-Violating and Birefringent Vacuum}
\end{center}

\vspace{1ex}

\begin{center}
{\large Brett Altschul\footnote{{\tt baltschu@indiana.edu}}}

\vspace{5mm}
{\sl Department of Physics} \\
{\sl Indiana University} \\
{\sl Bloomington, IN 47405 USA} \\

\end{center}

\vspace{2.5ex}

\medskip

\centerline {\bf Abstract}

\bigskip

We calculate the emission spectrum for vacuum Cerenkov radiation in
Lorentz-viola\-ting extensions of electrodynamics. We develop an approach
that works equally well if the presence or the absence of birefringence.
In addition to confirming earlier work, we present the first calculation
relevant to Cerenkov radiation in the presence of a birefringent photon $k_{F}$ 
term, calculating the lower-energy part of the spectrum for that case.

\bigskip

\end{titlepage}

\newpage

\section{Introduction}

In the past decade, a great deal of interest has developed in the
possibility that Lorentz and
CPT symmetries might not be exact in nature. If any violations of these important
symmetries were discovered, they would be of tremendous importance.
The form of the violations could potentially tell us great deal about the
new physics of the Planck scale. In fact, a number of
candidate theories of quantum gravity suggest the possibility of Lorentz symmetry
breaking in certain regimes. For example, Lorentz violation could arise
spontaneously in
string theory~\cite{ref-kost18,ref-kost19} or elsewhere~\cite{ref-altschul7}.
There could also be Lorentz-violating 
physics in loop quantum gravity~\cite{ref-gambini,ref-alfaro} and
non-commutative
geometry~\cite{ref-mocioiu,ref-carroll3} theories, or Lorentz violation through
spacetime-varying couplings~\cite{ref-kost20}, or
anomalous breaking of Lorentz and CPT symmetries~\cite{ref-klinkhamer}
in certain spacetimes.

Over the years, there have been many sensitive experimental tests of Lorentz
symmetry. Modern tests of this type have included studies of matter-antimatter
asymmetries for
trapped charged particles~\cite{ref-bluhm1,ref-bluhm2,ref-gabirelse,
ref-dehmelt1} and bound state systems~\cite{ref-bluhm3,ref-phillips},
determinations of muon properties~\cite{ref-kost8,ref-hughes}, analyses of
the behavior of spin-polarized matter~\cite{ref-kost9,ref-heckel2},
frequency standard comparisons~\cite{ref-berglund,ref-kost6,ref-bear,ref-wolf},
Michelson-Morley experiments with cryogenic resonators~\cite{ref-antonini,
ref-stanwix,ref-herrmann}, Doppler effect measurements~\cite{ref-saathoff,ref-lane1},
measurements of neutral meson
oscillations~\cite{ref-kost10,ref-kost7,ref-hsiung,ref-abe,ref-link,ref-aubert},
polarization measurements on the light from distant galaxies~\cite{ref-carroll1,
ref-carroll2,ref-kost11,ref-kost21}, analyses of the spectra of energetic
astrophysical sources~\cite{ref-jacobson1,ref-altschul6},
and others. There is a well-developed effective
field theory framework, the standard model extension (SME), which parameterizes
possible Lorentz violations in a local quantum field
theory~\cite{ref-kost1,ref-kost2} and also in the gravity sector~\cite{ref-kost12}.

The general SME has an infinite number of parameters, since it includes
nonrenormalizable operators of arbitrarily high dimensions.
Practically, it is usually more useful to restrict attention to a finite
subset of these operators. The most commonly considered subset is the minimal
SME. This includes operators which are superficially renormalizable (that is, of
dimension two, three, or four) and invariant under the standard model's
$SU(3)_{c}\times
SU(2)_{L}\times U(1)_{Y}$ gauge group. The minimal SME describes the forms of
Lorentz violation that should be most important at lower energies.
We shall only consider minimal SME operators
in this paper, although higher-dimension operators could still have distinct and
potentially quite interesting effects on the processes that we are interested in.
We shall also specialize to Lorentz violations that are entirely in the
electromagnetic sector, so that the matter sector is conventional.

Lorentz violating field theories are extremely interesting theoretically, since
they possess many new features that are absent in Lorentz-invariant models.
Processes that are kinematically forbidden when Lorentz symmetry is exact may
become allowed when this symmetry is weakly broken. One especially interesting
process is vacuum Cerenkov radiation, $e^{-}\rightarrow e^{-}+\gamma$. This is the
analogue of ordinary Cerenkov radiation in matter, and the threshold conditions
are similar.  An electron (or other charged particle) can emit low-energy Cerenkov
photons when the electron's velocity exceeds the photons' phase speed. (When the
photon energy becomes large enough that recoil effects are important, the threshold
conditions becomes more complicated. This is natural, because the crucial
quantity---the electron's velocity---does not remain constant through the duration of
the emission process. What must exceed the phase speed of light in this case is
the electron's average velocity during the emission process---averaged over the
region of momentum space between the initial and final values of the electron
momentum.)

The problem of vacuum Cerenkov radiation may be approached from several angles.
There are a number of different operators in the SME photon sector that could
give rise to this kind of process. How Cerenkov radiation works in the presence of
a birefringent Chern-Simons term has been analyzed in detail, using both
macroscopic techniques~\cite{ref-lehnert1,ref-lehnert2} and the microscopic
language of Feynman diagrams~\cite{ref-kaufhold}. Terms that do not induce
birefringence have also been considered~\cite{ref-altschul9}. However, there are
still ten coefficients in just the minimal SME photon sector whose effects have
not yet been considered in this context.

Our goal in this paper is to develop a technique that will allow us to
study the spectrum of
vacuum Cerenkov radiation in modified electrodynamic theories. This will
enable us to fill in some of the gaps in our knowledge of how the various
minimal SME terms impact the Cerenkov process.
However, there are some questions that we shall not be able to answer using this
method.
In order to make our calculations tractable, we must make some general simplifying
assumptions. However, all these assumptions are quite reasonable physically,
because we know that any deviations from conventional electrodynamics (whether
Lorentz violating or otherwise) must be very small at observable photon energies.
Our method will be macroscopic, relying on methods qualitatively similar to
those used in the calculation of ordinary Cerenkov radiation in dielectric materials.
We shall also neglect any recoil effects; because of this, and because of the
particular minimal SME operators that we are considering, the threshold condition
takes its simplest form.

There are three important aspects of vacuum Cerenkov radiation that distinguish it
from textbook Cerenkov radiation. These are dispersion, birefringence, and direction
dependence. Not all the theories we shall consider have all of these properties, but
understanding each of them will be important to a complete understanding of vacuum
Cerenkov radiation. Of course, dispersion exists in real materials as well as
Lorentz-violating vacua; the index of refraction will always be a function of
frequency. The other two effects, which involve direction- and polarization-dependent
speeds of light, are also seen in certain asymmetric crystals; however, they will
be much more important and generally more complicated in Lorentz-violating
field theories. Physical vacuum
birefringence at observable wavelengths is strongly constrained by
astrophysical experiments, but the property is still of significant theoretical
interest.

As an input to our Cerenkov radiation calculations, we shall require the dispersion
relations for the propagating modes of the electromagnetic fields. In the
birefringent case, there will be two separate dispersion relations for each wave
vector, which we shall denote by $\omega_{(i)}\!\left(\vec{k}\right)$, corresponding
to the polarization vectors $\hat{\epsilon}_{(i)}$. In general, the polarization
structure
of the normal modes of propagation will also depend on $\vec{k}$, and this dependence
may be either on the magnitude $k=\left|\vec{k}\right|$ or on the direction
$\hat{k}=\vec{k}/k$ (or potentially both). In situations
without birefringence, the choice of polarization basis is unimportant, because
all polarizations possess the same phase speed. In all cases we are considering,
Cerenkov radiation is
possible if the phase speed for at least one low-frequency mode of the
electromagnetic field is less than 1. 

We shall restrict our attention to the most physically relevant case, that in which
the deviation from 1 of the vacuum speed of light is small, $1-\omega\!\left(\vec{k}
\right)\!/k\ll 1$. This excludes some regimes of potential theoretical interest, but
it covers any region for which there is a reasonable possibility of actually
observing the Cerenkov radiation. Many of the excluded regions are also bedeviled by
problems with stability or causality~\cite{ref-kost3}, which could prevent us from
deriving meaningful results in any case.
We shall also assume that the deviation of the dispersion
relation from its conventional form is a slowly
varying function of $\vec{k}$---so that $\left|\vec{\nabla}_{\vec{k}}\left[\omega\!
\left(\vec{k}\right)\!/k\right]\right|\ll 1/k$.
Finally, we shall only consider linear modifications of
electrodynamics and theories with conventional source terms.

In general, with possible Lorentz and CPT violations, the electromagnetic field
of a propagating wave may not be transverse. This, however, is not an important
effect in the regime we are considering, in which the deviations from conventional
electrodynamics are small. To see why this effect is of secondary importance, we
may consider the following dichotomy: a change in the dispersion relation of the
electromagnetic waves without a change in the polarization structure can lead to
Cerenkov radiation; however a change in the polarization states without a change in
the dispersion relation cannot. Modifications to the phase speed are therefore more
important. The existence of non-transverse propagating waves will only result in
higher-order corrections to the effect we are interested in.
We shall therefore neglect any changes that the new physics may make to the
space of physical polarizations of the radiation field, and we assume that the
normal mode polarization vectors
$\hat{\epsilon}_{(1)}\!\left(\vec{k}\right)$ and $\hat{\epsilon}_{(2)}\!\left(\vec{k}
\right)$
span the transverse subspace with $\hat{\epsilon}_{(i)}\!\left(\vec{k}\right)\cdot
\vec{k}=0$.

Because the modified electrodynamical theories we shall be considering are linear, we
may work with each polarization mode separately. Because the sources of the field
are not modified, the crucial question for each mode of the field is how much
Cerenkov radiation a moving charge will emit with that wave vector and polarization,
and this question may be answered by relatively conventional means. We need only
calculate how much radiation would be emitted in that particular mode in an
ordinary Cerenkov process, in a medium with the right dielectric constant to
give the mode we are interested in the correct phase speed.

Our physically motivated approximations will also allow other simplifications.
The smallness of the deviation of the phase speed from 1 ensures that the
(appropriately generalized) Mach cone will always be very broad. Cerenkov photons
must be emitted in directions close to the direction $\hat{v}$ of the charge's
motion. Moreover, there will only be emission if the charge's speed is close to 1.

With our conventional matter sector, 1 is the maximum achievable velocity for a
moving charge; however, many of the expressions we shall derive would apply equally
well to theories with Lorentz violation in the matter sector and speeds $v>1$
allowed. However, we shall assume $v<1$, because it simplifies the accounting of
which modes of the electromagnetic field contain vacuum Cerenkov radiation.
In fact, it is not always a well posed question which sector actually
contains a Lorentz violation;
some forms of matter-sector Lorentz violation can be defined away,
a change of coordinates moving the Lorentz violations into the gauge sector without
changing the physics.

Since our study of the Cerenkov spectrum will be performed on a mode by mode basis,
there are some interesting features on vacuum Cerenkov radiation which it is not
possible to study with these techniques. It can be difficult or impossible to
calculate the back-reaction on the moving charge, because this quantity depends on
the total emission in all modes of the electromagnetic field. This limits our
ability to evaluate of the high-energy part of the spectrum, where recoil is an
important effect. It also limits us to considering moving charges with velocities
significantly above the Cerenkov threshold; if the velocity were too close to the
threshold, the recoil accompanying the first emitted photons could push the charge's
velocity back below threshold, fundamentally altering the character of the process.
Furthermore, for some modes, the
condition that the deviation of the phase speed from 1 be small may not be
met, and for higher-energy modes, new physics may come into play. Questions
related to the overall stability of the process may also be difficult to answer,
although they have already been considered, using a different technique, for the
case which poses the most interesting questions in this regard; in the
presence of a Lorentz-violating Chern-Simons term, there is no vacuum Cerenkov
emission if and only if in the rest frame of the moving charge the
energetic stability of the electromagnetic sector is
manifest~\cite{ref-lehnert1}. We shall also
neglect consideration of any finite-duration effects, treating the Cerenkov radiation
as a completely
steady state process. Thus, some interesting phenomena that arise in real, finite
period Cerenkov processes (such as diffraction) will not have any analogues in our
analysis.

This paper is organized as follows. In section~\ref{sec-models}, we introduce a
number of interesting theories, most of them Lorentz violating, in which vacuum
Cerenkov radiation could be possible. In section~\ref{sec-tech}, we show how to
generalize the usual equations governing Cerenkov radiation, and in
section~\ref{sec-appl}, we apply these generalizations to the specific models
introduced in section~\ref{sec-models}. Section~\ref{sec-concl} presents our
conclusions and outlook.

\section{Modified Electrodynamic Models}
\label{sec-models}

The free Lagrange density for the
electromagnetic sector, without any of the modifications that would make
vacuum Cerenkov radiation possible, is
\begin{equation}
{\cal L}_{0}=-\frac{1}{4}F^{\mu\nu}F_{\mu\nu}-j^{\mu}A_{\mu}.
\end{equation}
For our purposes here, the source term may be taken to be externally specified,
corresponding to a point charge $e$ moving with velocity $\vec{v}$ well above the
Cerenkov threshold.

In this section, we shall only be concerned with the free propagation of
electromagnetic waves in vacuum. This restricts us to considering modifications
of the Lagrangian which are bilinear in the electromagnetic field. We shall also
consider only those operators which are superficially renormalizable---that is,
operators of dimensions two, three, and four. We shall not insist on gauge
invariance for the dimension two operators, since there are potentially interesting
physics associated with photon mass terms. However, we shall only consider
higher-dimension operators that are gauge invariant, at least at the level of the
action.

We shall consider modifications to ${\cal L}_{0}$ one at a time.
Within the minimal SME, there are two types of gauge-invariant,
renormalizable Lorentz-violating
coefficients in the purely electromagnetic sector. (Since we are considering the
Cerenkov response to an externally prescribed charge density, we shall not consider
Lorentz violation in the matter sector, although Lorentz violations can have
important effects on how real particles move.) These are the CPT-odd Chern-Simons
term
\begin{equation}
{\cal L}_{AF}=\frac{1}{2}k_{AF}^{\mu}\epsilon_{\mu\nu\rho\sigma}F^{\nu\rho}
A^{\sigma}
\end{equation}
and the CPT-even term
\begin{equation}
{\cal L}_{F}=-\frac{1}{4}k_{F}^{\mu\nu\rho\sigma}F_{\mu\nu}F_{\rho\sigma}.
\end{equation}
The four-index tensor $k_{F}$ has the symmetries of the Riemann tensor and a
vanishing double trace, leaving it with nineteen independent coefficients.
We shall consider the $k_{AF}$ and $k_{F}$ separately, because doing so will make the
analysis much more elegant and intuitive. However, there would no impediment in
principle to doing our calculations in the presence of both $k_{AF}$ and $k_{F}$,
for which case the dispersion relations and the correct
techniques for identifying the elliptically polarized
normal modes of propagation are known~\cite{ref-kost2}.

The Chern-Simons term ${\cal L}_{AF}$ (which is gauge invariant up to a total
derivative) gives different dispersion relations for right-
and left-handed electromagnetic waves. The positive- and negative-helicity waves
have frequencies~\cite{ref-carroll1}
\begin{equation}
\label{eq-kAFdisp}
\omega_{\pm}^{2}=k^{2}\pm \frac{k_{AF}^{0}k-\left|
\vec{k}_{AF}\right|\omega_{\pm}\cos\theta_{AF}}{\sqrt{\frac{1}{4}-\frac{\vec{k}_{AF}
^{2}\sin^{2}\theta_{AF}}{\omega_{\pm}^{2}-k^{2}}}},
\end{equation}
where $\theta_{AF}$ is the angle between $\vec{k}$ and $\vec{k}_{AF}$.
Equation (\ref{eq-kAFdisp}) is not a closed form expression for $\omega_{\pm}\!
\left(\vec{k}\right)$; however, expanding it to leading order in the Lorentz
violation, we get
\begin{equation}
\label{eq-kAFappdisp}
\omega_{\pm}\approx k\pm\left(k_{AF}^{0}-\left|\vec{k}_{AF}\right|\cos
\theta_{AF}\right).
\end{equation}
This covers the range of $k$
that we are interested in. However, it should be noted that
at small wave numbers, the dispersion relation becomes problematical. If only
$k_{AF}^{0}$ is nonzero, then we have $\omega_{\pm}^{2}=k(k\pm 2k_{AF}^{0})$, so
that $\omega$ may become imaginary. There are then runaway solutions, which can be
fixed only by allowing for acausal signal propagation. It is not clear whether the
theory can be physically meaningful in this regime, especially when considering an
unconventional radiation process like Cerenkov radiation. The previous macroscopic
analyses of vacuum Cerenkov radiation in the presence of $k_{AF}$ have focused
more on the spacelike case, in which such instabilities do not
occur.

With $k_{F}$ only, the dispersion relations are determined by the matrix equation
\begin{equation}
\label{eq-eig}
\left[\delta_{jk}p^{2}+p_{j}p_{k}+2\left(k_{F}\right)_{j\alpha\beta k}p^{\alpha}
p^{\beta}\right]\left[\hat{\epsilon}_{(i)}\right]_{k}=0,
\end{equation}
where $p^{\mu}=\left(\omega,\vec{k}\right)$ is the photon four-momentum.
In this case, the phase speed and
polarization structure are independent of the wave number,
although they do depend on the propagation direction.
To leading order in $k_{F}$, the polarizations are transverse and orthogonal,
and the frequencies are~\cite{ref-kost16}
\begin{equation}
\omega_{\pm}\approx\left[1+\rho\!\left(\hat{k}\right)\pm\sigma\!\left(\hat{k}\right)
\right]k,
\end{equation}
where $\rho\!\left(\hat{k}\right)=-\frac{1}{2}\tilde{k}^{\alpha}\,_{\alpha}$, and
$\sigma^{2}\!\left(\hat{k}\right)=\frac{1}{2}\tilde{k}^{\alpha\beta}\tilde{k}
_{\alpha\beta}-\rho^{2}\!\left(\hat{k}\right)$, with $\tilde{k}^{\alpha\beta}=
k_{F}^{\alpha\mu\beta\nu}\hat{p}_{\mu}\hat{p}_{\nu}$ and $\hat{p}^{\mu}=
\left(1,\hat{k}\right)$.
Since the $k_{F}$ term is dimensionless, there is no dispersion in the photon
spectrum; the phase speed depends only on $\hat{k}$. Moreover,
there is no birefringence if $\sigma=0$, which was the case considered
in~\cite{ref-altschul9}.

Using the explicit leading order expression for the dispersion relation, it is
possible to recast the eigenvector condition (\ref{eq-eig}) as
\begin{equation}
\left[2(\rho\pm\sigma)\delta_{jk}-\hat{k}_{j}\hat{k}_{k}+2\tilde{k}_{jk}\right]
\left[\hat{\epsilon}_{(\pm)}\right]_{k}=0.
\end{equation}
In a primed frame where a photon's energy-momentum is $\hat{p}'^{\mu}=\left(1,
\hat{e}_{3}\right)$,
the polarization vectors are
\begin{equation}
\label{eq-epspm}
\hat{\epsilon}'_{(\pm)}\propto(\sin\xi,\pm1-\cos\xi,0),
\end{equation}
where $\sigma\sin\xi=\tilde{k}'_{12}$ and $\sigma\cos\xi=\frac{1}{2}\left(\tilde{k}'
_{11}-\tilde{k}'_{22}\right)$. In these coordinates, $\hat{\epsilon}_{(+)}$ makes
an angle $\xi/2$ with the $x'$-axis.

Another class of possibly Lorentz-violating models with photon speeds less than 1
may also be considered. These are models which break gauge invariance. The
Lagrange density
\begin{equation}
{\cal L}_{M}=M^{\mu\nu}A_{\mu}A_{\nu}
\end{equation}
has a generalized photon
mass term. In the presence of ${\cal L}_{M}$, there are generally three propagating
modes of the electromagnetic field. However, we shall neglect the novel longitudinal
mode, because if the breaking of gauge invariance is weak, this mode will be
correspondingly weakly coupled to charges.

Of interest is obviously the Lorentz-invariant Proca theory, with $M^{\mu\nu}=
\frac{1}{2}g^{\mu\nu}m^{2}$. Other Lorentz-violating versions have also aroused some
recent interest~\cite{ref-gabadadze,ref-dvali,ref-altschul10}.
In all the cases that have been considered, there exists a frame in
which $M^{\mu}\,_{\nu}$ is diagonal, with non-negative eigenvalues, at most
one of which is different from the others. These models are never birefringent.
If $M^{0}\,_{0}$ vanishes, but the $M^{j}\,_{k}=\frac{1}{2}m_{1}^{2}\delta^{j}\,_{k}$
are nonvanishing, then the theory contains only two propagating modes;
while if $M^{0}\,_{0}$ is finite as well, then there is also a propagating
longitudinal mode. However, the dispersion relation for the transverse modes is
always
\begin{equation}
\omega^{2}=k^{2}+m_{1}^{2}.
\end{equation}

If one of the spatial elements on the diagonal of $M^{\mu}\,_{\nu}$ differs from the
others, then the situation is more complex. The basis of propagation states is
not orthogonal; however, the only mode with an unconventional dispersion relation is
again essentially longitudinally polarized. (As the Lorentz violation is gets
smaller, the
associated polarization vector moves closer to $\hat{k}$.) The net result is that
the transverse modes propagate at the same rate, and this rate is again independent
of the propagation direction.

So for all the $M^{\mu}\,_{\nu}$ of interest, the transverse modes have the same
type of dispersion relation. Unfortunately, while the group velocity for
this dispersion relation is always less than 1, the phase velocity is the
reciprocal of the group velocity and is hence always greater than 1. There is
thus no vacuum Cerenkov radiation in these theories, and we shall not consider them
any further.

\section{Calculational Techniques}
\label{sec-tech}

We must now generalize the usual techniques used to calculate rates of
Cerenkov emission to cover the vacuum cases we are interested in.
For a charge $e$ moving with velocity $\vec{v}$, subject to $v<1$, there may be
Cerenkov emission if $v$ is greater than the phase speed of light in some direction.
In a conventional dielectric material, where the index of refraction is direction-
and polarization-independent and
constant (or only slowly varying) as a function of wave number, there is a sharp
Mach cone, with a discontinuity (or near discontinuity)
across it. The moving charge emits photons, and
the cone represents the signal front for their propagation.
The Cerenkov angle $\theta_{C}$ is the angle between the direction $\hat{v}$ and
the propagation of the emitted photons. With dispersion, this angle becomes
frequency dependent. In an anisotropic vacuum, it will also generally depend on
the azimuthal angle around $\hat{v}$, and with birefringence, it will depend on the
the polarization as well. In the birefringent case, there may exist Cerenkov
emission for only one of the polarizations corresponding to a given $\vec{k}$, since
the phase speeds for the two polarizations are not the same.
The opening angle of the Mach cone is $\frac{\pi}{2}-\theta_{C}$.

If there is significant
dispersion, there will not generally be a Mach cone defined by a sharp shock front.
However, the radiation at a given fixed frequency will all be located on a cone,
although with Lorentz violation, that cone need not be right angled or circular.

In the ordinary case, $\theta_{C}$ is determined by the coherence condition,
\begin{equation}
\label{eq-oldcoh}
\cos\theta_{C}=\frac{1}{vn}.
\end{equation}
This can be thought of as simply the angle at which the radiation in the far field
does not interfere with itself destructively, so that there is a net nonzero
Poynting flux. (Since interference between waves emitted by the moving charge
at different points in time is the crucial effect, the derivation of the coherence
condition requires that the velocity not change appreciably during the time
period under consideration.)
The same constructive interference is still necessary in the Lorentz-violating case
if there is to be a net outflow of radiation.
So an analogous condition will govern the direction
in which radiation is emitted in the more complicated cases under consideration here;
however, the analog of (\ref{eq-oldcoh}) is no longer a
straightforward expression for
$\theta_{C}$ as a function of the other parameters. The effective index of refraction
can be a function of direction and hence of $\theta_{C}$. To generalize
(\ref{eq-oldcoh}), we note (omitting any dependences on polarization) that
$\cos\theta_{C}=\hat{v}\cdot\hat{k}$, and $n$ generalizes to $k/\omega\!\left(\vec{k}
\right)$. A simple rearrangement then yields the generalized coherence condition
\begin{equation}
\label{eq-coh}
\vec{v}\cdot\vec{k}=\omega\!\left(\vec{k}\right).
\end{equation}

For arbitrary anisotropic dispersion relations, (\ref{eq-coh}) may be difficult
to solve. However, when the effects of new physics are small, we may solve
(\ref{eq-coh}) perturbatively. When $1-\omega\!\left(\vec{k}\right)\!/k\ll 1$, the
angle $\theta_{C}$ will be small. The photons are all emitted in directions
$\hat{k}$ very close to $\hat{v}$, so to leading order we may calculate $\theta_{C}$
by approximating $\vec{k}$ by $k\hat{v}$ on the right-hand side of (\ref{eq-coh}).
This gives
\begin{eqnarray}
\cos\theta_{C}\approx1-\frac{\theta_{C}^{2}}{2}& \approx & \frac{\omega\left(k\hat{v}
\right)}{kv} \\
\label{eq-thetaC}
\theta_{C}^{2} & \approx & 2\left[1-\omega\left(k\hat{v}\right)/kv\right].
\end{eqnarray}
Cerenkov radiation is emitted if $\theta_{C}^{2}>0$.
At this level of approximation, the Mach cone is right-angled and circular. Any
obliquities are higher-order effects. However, the opening angle of the cone will
vary with the direction $\hat{v}$ of the charge's movement. We shall make
extensive further use of the approximation $\vec{k}\approx k\hat{v}$ in obtaining
other leading order results; this amounts to approximating the Mach cone
as being a flat planar wave front.

In the absence of birefringence, (\ref{eq-thetaC}) is all that is needed to determine
the
leading-order character of the radiation. Because the theory is linear, we may look
at the electromagnetic field one mode at a time. In a momentum-space neighborhood
of any given mode, the theory looks like conventional electrodynamics but
with a speed of light $\omega\!\left(\vec{k}\right)\!/k\equiv 1/n\!\left(\vec{k}
\right)$ different from 1. The effects of dispersion are suppressed, because
$\omega\!\left(\vec{k}\right)\!/k$ is slowly varying.

However, we must still deal with the birefringent case.
Ordinary Cerenkov radiation is linearly polarized in the plane defined by $\hat{v}$
and $\hat{k}$; we shall denote the corresponding polarization vector by
$\hat{\epsilon}_{(0)}\!
\left(\vec{k}\right)$. Obviously, birefringence will change the polarization of
the emitted radiation. However, the changes are
really quite simple, again because we can look at the theory one mode of the field
at a time. If the coherence condition is
satisfied for a given mode of the field, with wave vector $\vec{k}$ and
polarization $\hat{\epsilon}_{(i)}$, what happens in all the orthogonal modes is
unimportant. In particular, the emission in the mode we are interested in is
exactly the same as in any other theory with the same $\vec{v}$ and $n\!\left[\vec{k}
,\hat{\epsilon}_{(i)}\right]$. So we can imagine replacing the theory by one
with a constant, polarization-independent index of refraction $\tilde{n}=n\!\left[
\vec{k},\hat{\epsilon}_{(i)}\right]$. Calculating the emitted power in the mode of
interest is then trivial. It is just the total power emitted in the conventional
theory with $\tilde{n}$, times the squared overlap between the conventional linear
polarization mode and the mode we are studying, or
just $\left|\hat{\epsilon}_{(i)}\cdot\hat{\epsilon}_{(0)}\right|^{2}$. This
technique for determining the radiated power is valid even beyond the leading order
approximation.

The power emitted per unit frequency in ordinary Cerenkov radiation is
$P(\omega)=\frac{e^{2}}{4\pi}
\sin^{2}\theta_{C}\omega$. The generalizations required by the presence
of Lorentz violations and dispersion are minor, at least at leading order.
$\theta_{C}^{2}$ must be determined from
(\ref{eq-thetaC}), and we must include the polarization overlap factor, which
adds the only real complication. At leading order, the Mach cone is right and
circular, so $\theta_{C}$ does not depend on the azimuthal angle $\phi$. Nor,
at leading order, do the normal mode polarization vectors $\hat{\epsilon}_{(i)}$
depend on $\phi$---and for the same reason, since we can approximate the
$\hat{\epsilon}_{(i)}\!\left(\vec{k}\right)$ by $\hat{\epsilon}_{(i)}(k\hat{v})$.
However, the polarization of ordinary Cerenkov radiation does depend strongly on
$\phi$. In the same leading-order approximation we have been using, the polarization
vector $\hat{\epsilon}_{(0)}$ is
$\hat{\rho}\equiv\hat{\phi}\times\hat{v}$. So the emitted power may depend on
$\phi$. We must therefore express the power per unit frequency per unit of
azimuthal angle for a given polarization mode; this is
\begin{equation}
\label{eq-power}
P_{(i)}(\omega,\phi)
\approx\frac{e^{2}}{(2\pi)^{2}}\left|\hat{\epsilon}_{(i)}\cdot\hat{\rho}
\right|^{2}\left[1-\omega_{(i)}\!\left(k\hat{v}\right)/kv\right]\omega_{(i)}\!
\left(k\hat{v}\right).
\end{equation}
[At this level of approximation, we may also replace the terminal $\omega_{(i)}$ by
$k$.]
The average of $\left|\hat{\epsilon}_{(i)}\cdot\hat{\rho}\right|^{2}$ over all
$\phi$ is always $\frac{1}{2}$. In a theory without birefringence, all the factors in
(\ref{eq-power})
except $\left|\hat{\epsilon}_{(i)}\cdot\hat{\rho}\right|^{2}$ are equal for the two
polarizations, and the total emitted power, summed over both polarizations, is
independent of $\phi$, as we might expect.

To leading order, we need not distinguish whether our expression for the power
is the power per unit frequency or per unit
wave number, nor need we worry about how the energy-momentum tensor
is modified by the new physics; these effects are only important beyond leading
order. So in (\ref{eq-power}), we have a simple expression for the leading-order
Cerenkov spectrum in any normal mode of the radiation field that satisfies our
assumptions.

\section{Application to Specific Models}
\label{sec-appl}

We may now apply our method to the models discussed in
section~\ref{sec-models}.
Considering the $k_{AF}$ model first, we note that, according to the leading order
expression (\ref{eq-kAFappdisp}) for the dispersion relation, exactly one mode
of the field is superluminal and one subluminal for each value of $\vec{k}$.
Furthermore, since the normal modes of propagation are always circularly polarized
waves, the overlap expression $\left|\hat{\epsilon}_{(i)}\cdot\hat{\epsilon}_{(0)}
\right|^{2}$ is always exactly equal to $\frac{1}{2}$. The Cerenkov angle is also
easy to calculate at leading order. For whichever polarization is moving more
slowly than $v$, it is
\begin{equation}
\label{eq-thetaAF}
\theta_{C}^{2}\approx 2\left(1-\frac{1}{v}+\frac{\left|k_{AF}^{0}-\vec{k}_{AF}\cdot
\hat{v}\right|}{kv}\right).
\end{equation}
Since this depends on $k$, there is not a sharply defined Mach cone.
The sign of the expression inside the absolute value determines which polarization
this represents. The helicity of the emitted photons is $-{\rm sgn}\left(
k_{AF}^{0}-\vec{k}_{AF}\cdot\hat{v}\right)$; this agrees with the results
in~\cite{ref-lehnert2} for emission close to the direction of $\vec{v}$.

The total power emitted per unit frequency is
\begin{equation}
P(\omega)\approx\frac{e^{2}}{4\pi}\left(1-\frac{1}{v}+\frac{\left|k_{AF}^{0}-
\vec{k}_{AF}\cdot\hat{v}\right|}{kv}\right)k,
\end{equation}
and this is emitted in an azimuthally symmetric pattern around $\hat{v}$.
It is clear in this case that the dispersion will cut off the Cerenkov spectrum
at high energies, because the absolutely value term in (\ref{eq-thetaAF}) is
divided by $k$. For large wave numbers
$k>\left|k_{AF}^{0}-\vec{k}_{AF}\cdot\hat{v}\right|/(1-v)$,
there is no emission, and this ensures that there is no ultraviolet divergence
in the total power.

\begin{figure}[t]
\epsfxsize=3in
\begin{center}
\leavevmode
\epsfbox[300 350 550 600]{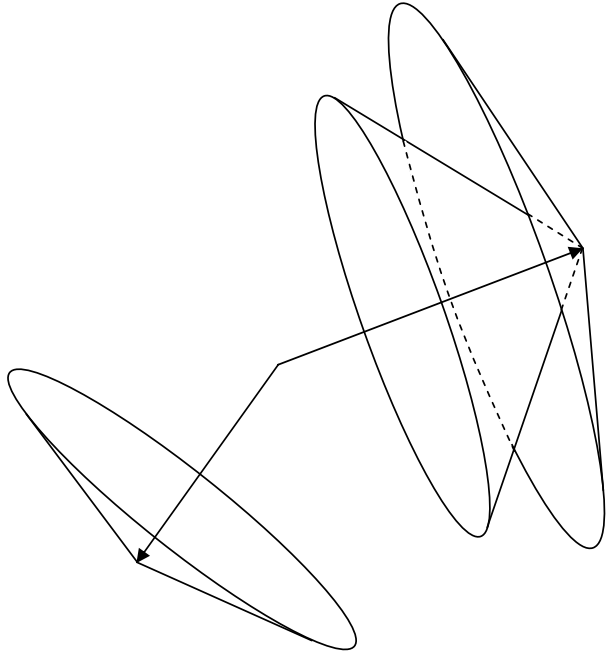}
\caption{Possible shapes of the Mach cones corresponding to charges moving in two
different directions. In one direction, two Mach cones are possible, but in another
direction only one. In the leading order approximation, the cones are right-angled
and circular, although the cones shown here are exaggeratedly narrow.
\label{fig-cones}}
\end{center}
\end{figure}

For the theory with $k_{F}$, the results are equally straightforward. In this case,
if $\rho(\hat{v})<-|\sigma(\hat{v})|$, there can be Cerenkov radiation in both
polarization modes. However, unless $\sigma(\hat{v})=0$, the two polarizations will
have Mach cones of different width and different rates of emission. Indeed, the
Cerenkov angles are
\begin{equation}
\theta_{C}^{2}\approx
2\left[1-\frac{1}{v}-\frac{\rho(\hat{v})\pm\sigma(\hat{v})}{v}\right].
\end{equation}
Figure~\ref{fig-cones} shows the possible shapes of the Mach cones for two
different particle velocities $\vec{v}$. In one direction, there are two broad
cones, corresponding to two different polarization states, but in the second
direction, only a single Mach cone is possible.

\begin{figure}[t]
\epsfxsize=3in
\begin{center}
\leavevmode
\epsfbox[200 330 600 600]{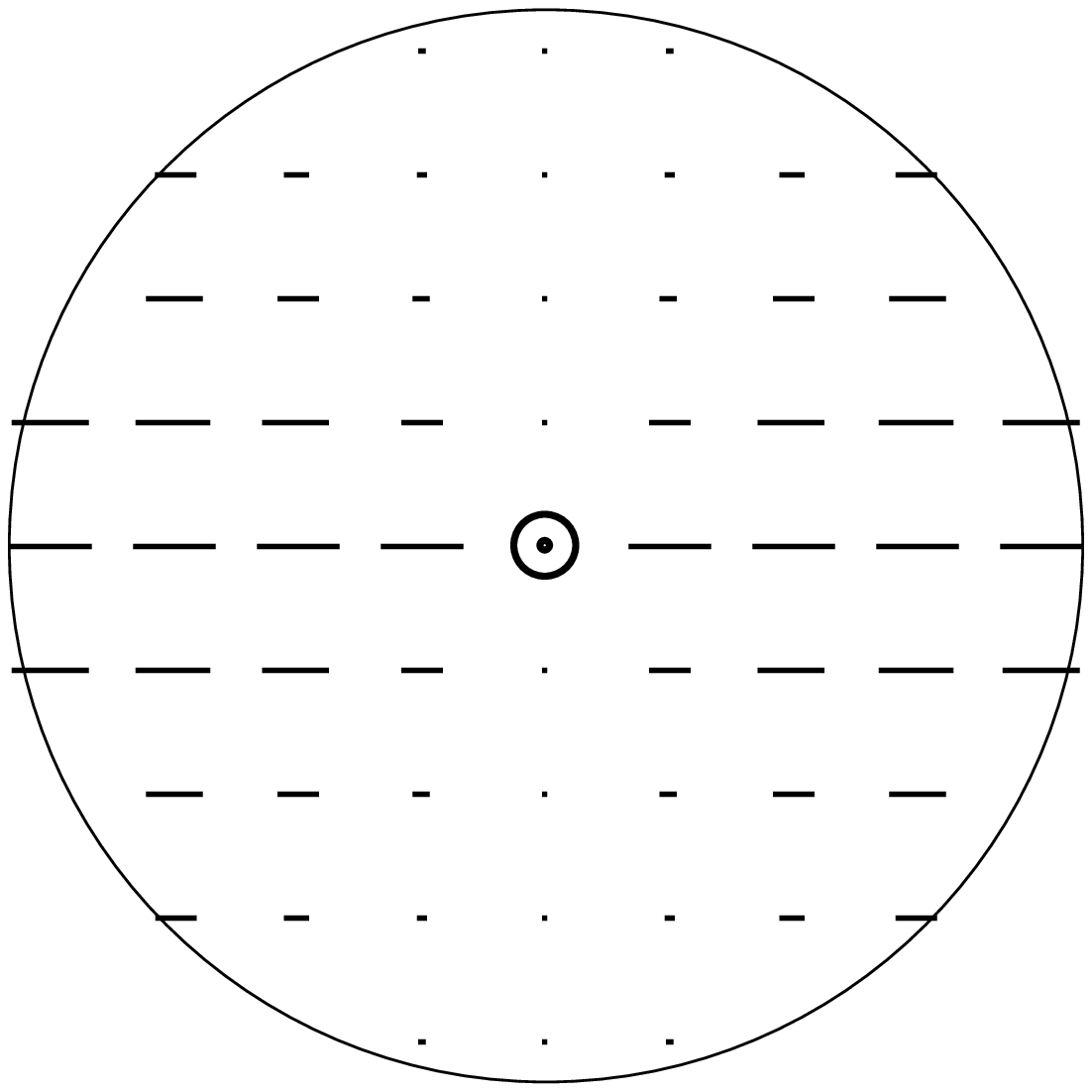}
\caption{Polarization of the
radiation on a single Mach cone in the birefringent $k_{F}$
theory. The cone is seen from above, with the charge moving out of the plane toward
the viewer. The lines indicate the polarization direction of the radiation at
various points on the cone's surface, with their lengths denoting the relative
intensities.
\label{fig-pol}}
\end{center}
\end{figure}

If we choose coordinates so that $\hat{v}=\hat{e}_{3}$, the polarization vectors
corresponding to the two cones are given by (\ref{eq-epspm}). The angular overlap
factors are then, using $\hat{\phi}\times\hat{v}=(\cos\phi,\sin\phi,0)$,
\begin{eqnarray}
\left|\hat{\epsilon}_{(+)}\cdot\hat{\epsilon}_{(0)}\right|^{2} & \approx
& \cos^{2}(\phi-\xi/2) \\
\left|\hat{\epsilon}_{(-)}\cdot\hat{\epsilon}_{(0)}\right|^{2} & \approx
& \sin^{2}(\phi-\xi/2).
\end{eqnarray}
If $\sigma(\hat{v})=0$, then the radiation intensity is independent of $\phi$;
moreover, the
Cerenkov angle---and thus the power emitted---agrees in this case with the results
calculated in~\cite{ref-altschul9} by somewhat different methods.
However, because there more generally is birefringence, there can be an angular
dependence in the total power. The two polarizations are emitted in two
perpendicularly-oriented dipole-like patterns. (These are not, however, dipole
radiation patters in the usual sense, since the waves are all directed into a
narrow angular range around the direction $\hat{v}$.) Figure~\ref{fig-pol} shows
the intensity and polarization on the surface of one of the two possible Mach cones.

The expressions for $\theta_{C}^{2}$ and the polarization overlap factors give
us the power emitted and its angular distribution,
\begin{eqnarray}
P_{(+)}(\omega,\phi) & \approx & \frac{e^{2}}{(2\pi)^{2}}
\left[1-\frac{1}{v}-\frac{\rho(\hat{v})+\sigma(\hat{v})}{v}\right]
\omega\cos^{2}(\phi-\xi/2) \\
P_{(-)}(\omega,\phi) & \approx & \frac{e^{2}}{(2\pi)^{2}}
\left[1-\frac{1}{v}-\frac{\rho(\hat{v})-\sigma(\hat{v})}{v}\right]
\omega\sin^{2}(\phi-\xi/2).
\end{eqnarray}
These results should be accurate
whenever our approximations are valid. However, with the $k_{F}$ term, there is
an obvious problem with the spectrum at large $k$. Since $\theta_{C}^{2}$ and
the polarization factors are independent of $k$, the spectrum appears to diverge
at high energies.

Some new effect, not considered here, must enter to cut off the spectrum. We have
neglected the recoil of the emitting particle. Including it ought to render the
whole expression finite; the charge will not radiate away more energy than it
possesses~\cite{ref-jacobson3}. So there is a natural cutoff at the energy scale
of the radiating particle itself. However, this is not necessarily the relevant
cutoff. New physics may enter at a scale lower than the energy scale of the particle,
and the scale of the new physics may represent the physically meaningful cutoff
scale.

These questions were previously discussed in~\cite{ref-altschul9}
and a partial solution put forward for the special case considered there. The
part of $k_{F}$ that does not cause birefringence mixes with Lorentz-violating
coefficients in the matter sector under renormalization~\cite{ref-kost4}. Although
the pure gauge
sector does not make reference to any mass scale, the matter sector will contain
massive charged particles, which will set a definite scale for the theory. This
implies that
new physics must enter at a scale $\Lambda\sim mk_{F}^{-1/2}$, where $m$ is the
lightest
charged particle mass; without the appearance of new physics, the theory will
exhibit pathological properties at high energies~\cite{ref-kost3}.
$\Lambda$ is essentially the largest scale at which new physics can enter, and
it is comparable to the threshold energy for the Cerenkov process.

The appearance of a high-energy cutoff comparable to the energy
threshold for vacuum Cerenkov is actually desirable, since it provides a
uniform ultraviolet regulator for the
total power emitted (although the
Cerenkov spectrum, even with the cutoff, still has a number of counterintuitive
properties~\cite{ref-altschul9}).
In this special case, any charge emitting vacuum Cerenkov of radiation must have an
energy comparable to or larger than $\Lambda$, so we expect $\Lambda$---not the
charge's energy---to be the most relevant cutoff scale. A charge whose energy is
far above the threshold level can emit radiation at a rate limited by $\Lambda$ for
an extended period, with the velocity decaying only comparatively slowly over this
time, so that recoil effects are unimportant.

However, the birefringent part of $k_{F}$ does
not mix with any other Lorentz-violating coefficients at leading order, and so an
electromagnetic theory containing only this form of Lorentz violation is equally
valid at all energy scales. It is obviously possible that new physics may cut off
this theory as well, but there is no indication of at what scale that cutoff
should come. Or it may be impossible to consider this theory without taking into
account the back-reaction on the charge, which loses momentum as it radiates.
Unlike the previous case, it is possible that recoil effects might provide the only
cutoff for the Cerenkov spectrum.
In any case,
the spectrum we have calculated should be a perfectly valid first approximation at
sufficiently low energies, but to understand the totality of the vacuum Cerenkov
process in the presence of this kind of Lorentz violation, a different approach
is required.

The new physics which should cut off the theory with a non-birefringent $k_{F}$ enter
the photon sector through radiative corrections, and it would be interesting to
consider how other loop effects might impact our results. One type of effect may be
of particular interest---photon splitting, $\gamma\rightarrow N\gamma$, which is
forbidden on shell in a gauge- and Lorentz-invariant theory. Photon splitting
amplitudes are generally nonzero in the presence of Lorentz violation,
however~\cite{ref-kost5}. An entirely speculative yet
interesting possibility in the context of vacuum
Cerenkov radiation is that the emitted photons may split into multiple collinear
photons, and the amplitude for this process may interfere destructively with the
primary Cerenkov amplitude. To determine whether this could actually occur would
require evaluation of the amplitude for many-photon Cerenkov emission, including its
phase, as well as the
photon splitting amplitude in the presence of a general $k_{F}$; and neither of these
amplitudes is known at present. A tricky balancing between terms at different orders
in $e^{2}$ would also be required in this scenario.
So whether photon splitting or other radiative
corrections have important impacts on the Cerenkov process is unknown.

\section{Conclusion}
\label{sec-concl}

In this paper, we have worked out a method for determining the spectrum of
vacuum Cerenkov radiation, working with the electromagnetic field mode by mode.
For our leading order results to be useful for a given mode, Lorentz violations must
affect the energy-momentum relation for the mode in question only slightly.
However, apart from that, the method is rather general, allowing us to treat the
$k_{AF}$ and $k_{F}$ terms, birefringent and not, on equal footing.
We have rederived previous results for a number of cases, and we have also provided
the first calculation of the lower-energy part of the Cerenkov spectrum in the
presence of a birefringent $k_{F}$.

However, there are many interesting questions that are still unanswered.
There are regions of the spectrum where our approximations are simply not
valid.  The small $k$ region in the $k_{AF}$ theory is outside the realm of our
approximations' validity, although it has been examined by other means. The large
$k$ domain in the presence of $k_{F}$  (where new physics may come into play to cut
off the Cerenkov process) raises further questions. Much more could also
be said about the back-reaction on the moving charge. One technique that might
make it easier to take this effect into account would be to calculate the Cerenkov
spectrum using Feynman diagrams, as was done for the $k_{AF}$ in~\cite{ref-kaufhold}.
Without understanding recoil effects, it may be impossible to calculate the
highest-energy portions of the Cerenkov spectrum or even the low-energy spectrum
for particles with energies very close to the threshold.
Because these kinds of deep questions
still exist, vacuum Cerenkov radiation remains a very interesting area in the
study of Lorentz violation.

\section*{Acknowledgments}
The author is grateful to V. A. Kosteleck\'{y} for helpful discussions.
This work is supported in part by funds provided by the U. S.
Department of Energy (D.O.E.) under cooperative research agreement
DE-FG02-91ER40661.

\end{document}